\documentclass{aa}
\usepackage{graphicx}
\usepackage{natbib}
\usepackage{txfonts}
\usepackage{amssymb}
\bibpunct{(}{)}{;}{a}{}{,} 
\begin{document}

   \title{Sizes of transition-region structures in coronal holes and in the quiet Sun}
   \author{H. Tian\inst{1,2}
           \and
           E. Marsch\inst{1}
           \and
           C.-Y. Tu\inst{1,2}
           \and
           L.-D. Xia\inst{3}
           \and
           J.-S. He\inst{2}
          }
   \offprints{H. Tian}
   \institute{Max-Planck-Institut f\"ur Sonnensystemforschung, Katlenburg-Lindau, Germany\\
              \email{tianhui924@163.com, marsch@mps.mpg.de}
             \and
             Department of Geophysics, Peking University, Beijing, China\\
             \email{chuanyitu@pku.edu.cn}
             \and
             Department of Space Science and Applied Physics, Shandong Univ. at Weihai, Weihai, Shandong, China\\
             \email{xld@ustc.edu.cn}
             }
   \date{}

\abstract
{}
{We study the height variations of the sizes of chromospheric and
transition-region features in a small coronal hole and the adjacent
quiet Sun, considering images of the intensity, Doppler shift, and
non-thermal motion of ultraviolet emission lines as measured by
SUMER (Solar Ultraviolet Measurements of Emitted Radiation),
together with the magnetic field as obtained by extrapolation from
photospheric magnetograms.}
{In order to estimate the characteristic sizes of the different
features present in the chromosphere and transition region, we have
calculated the autocorrelation function for the images as well as
the corresponding extrapolated magnetic field at different heights.
The Half Width at Half Maximum (HWHM) of the autocorrelation
function is considered to be the characteristic size of the feature
shown in the corresponding image.}
{Our results indicate that, in both the coronal hole and quiet Sun,
the HWHM of the intensity image is larger than that of the images of
Doppler-shift and non-thermal width at any given altitude. The HWHM
of the intensity image is smaller in the chromosphere than in the
transition region, where the sizes of intensity features of lines at
different temperatures are almost the same. But in the upper part of
the transition region, the intensity size increases more strongly
with temperature in the coronal hole than in the quiet Sun. We also
studied the height variations of the HWHM of the magnetic field
magnitude $B$ and its component $\left\vert B_z \right\vert$, and
found they are equal to each other at a certain height below 40 Mm
in the coronal hole. The height variations of the HWHM of
$\left\vert B_z/B\right\vert$ seem to be consistent with the
temperature variations of the intensity size.}
{Our results suggest that coronal loops are much lower, and magnetic
structures expand through the upper transition region and lower
corona much more strongly with height in the coronal hole than in
the quiet Sun.}


\titlerunning{Sizes of transition-region structures}
\authorrunning{H. Tian et al.}
\maketitle

\section{Introduction}

The solar transition region, where above the photosphere the
temperature increases rapidly and the density drops dramatically, is
a rather inhomogeneous and dynamic layer between chromosphere and
corona (for a review see \cite{Mariska1992}). The transition region
is believed to play an important role in the origin of the solar
wind and in coronal heating \citep{HasslerEtal1999, TuEtal2005a,
ChaeEtal1998, Hansteen1993}.

One of the most prominent features in the chromsphere and transition
region is the magnetic network \citep{Reeves1976}, which is the
upward extension of the supergranular boundaries above the photosphere.
The network manifests itself as bright lanes on the radiance
images of emission lines, and is characterized by clusters of
magnetic flux concentrations \citep{PatsourakosEtal1999}. The typical
network cell has a size of 20,000 to 30,000~km \citep{SimonLeighton1964},
and lasts for about 20 to 50 hours \citep{SimonLeighton1964, Schrijver1997, RajuEtal1998}.
It is suggested that part of the network flux opens into the corona
and has a funnel shape, while the rest of the network consists of a
dense population of low-lying loops with lengths less than $10^4$~km
\citep{DowdyEtal1986} and varying orientations. A funnel may be connected
to the solar wind or just be the foot of a large coronal loop \citep{Peter2001}.

Another prominent characteristic of the transition region is the
observed red shift of most transition region lines (see the reviews
by \cite{Mariska1992} and \cite{WilhelmEtal2007}). However, a net
blue shift of the upper transition region line
Ne~{\sc{viii}}~($\lambda770$) has been observed by SUMER (Solar
Ultraviolet Measurements by Emitted Radiation)
\citep{DammaschEtal1999}. The transition from redshift to blueshift
occurs at an electron temperature of about $5\times10^5$~K; the
detailed temperature variation of the Doppler shift can be found in
\cite{PeterJudge1999}, \cite{TeriacaEtal1999}, and
\cite{XiaEtal2004}.

Besides line intensity and Doppler shift, spectroscopic diagnostics
can also provide us with information on non-thermal motions of the
emitting ions. The observed non-thermal broadenings could correspond
to small-scale laminar flows, waves, or turbulence, each of which
may play a role in energy transport and coronal heating
\citep{ChaeEtal1998}. The non-thermal velocity at temperatures below
$2\times10^4$~K is less than 10~km~s$^{-1}$. On average, it
increases with temperature, reaches a peak value of 30~km~s$^{-1}$
around $3\times10^5$~K, and then decreases with temperature
\citep{ChaeEtal1998}. The average non-thermal velocities of
transition region lines are found to be larger in coronal holes than
in quiet Sun regions \citep{LemaireEtal1999}.

Investigating the temperature or height dependence of a specific
line feature such as radiance, Doppler shift and non-thermal width
in the solar atmosphere is very helpful in understanding the
structure of the solar atmosphere and the relevant physical
processes \citep{ChaeEtal1998}. Based on a two-dimensional
autocorrelation method, \cite{PatsourakosEtal1999} analyzed six
intensity images of transition region lines with different formation
temperatures, and thus studied the temperature variation of the
network-boundary thickness indicated by the HWHM of the
autocorrelation function (ACF) in a quiet Sun region. They concluded
that network boundaries have an almost constant size up to a
temperature of about $10^{5.4}$~K and then fan out rapidly at
coronal temperatures. \cite{GontikakisEtal2003} extended this work
by studying the sizes of structures in Doppler shift and width, but
only for three lines in a quiet Sun region. They found that the size
of bright radiance features is larger than that of the structures of
the dopplergram and Doppler width. They also found that the sizes of
C~{\sc{iv}} line are smaller than those of Si~{\sc{ii}}.
\cite{RavindraVenk2003} used the autocorrelation technique, as well
as the so-called structure function, to study the life time and
length scale of the network cells seen in images of He~{\sc{ii}}
(304 {\AA}) passband observed by the Extreme-ultraviolet Imaging
Telescope (EIT). The two methods were also applied to the
extrapolated magnetic field at different heights in their paper.

Here we will analyse three different data sets obtained by SUMER on
the Solar and Heliospheric Observatory (SOHO) and report the new
results on the sizes of chromospheric and transition-region
features. The autocorrelation technique was applied to images of
intensity, Doppler shift, and non-thermal width of several lines, as
well as the corresponding extrapolated magnetogram at different
heights. The HWHM of the autocorrelation function is considered to
be the characteristic size of the feature in the corresponding
image. We will investigate the height variations of the sizes of
different features, compare the differences found between coronal
hole and quiet Sun, and subsequently discuss the relevant physical
implications.

\section{Observations and data reduction}

Three data sets obtained by SUMER were included in this study. The
first one was taken from 18:00 on March 7 to 17:42 UTC of the next
day in 1997. During this period, SUMER observed an equatorial
coronal hole and the surrounding quiet Sun. More information about
this observation can be found in \cite{LemaireEtal1999},
\cite{AiouazEtal2005} and \cite{Aiouaz2008}, in which this data set
was also used.
We selected 6 emission lines including Ne~{\sc{viii}} (770.428
{\AA}), O~{\sc{v}} (760.43 {\AA}), O~{\sc{iv}} (787.70 {\AA}),
S~{\sc{v}} (786.47 {\AA}), N~{\sc{iv}} (765.15 {\AA}) and
N~{\sc{iii}} (764.36 {\AA}) for our study. We averaged the data over
4 exposures, which greatly improved the signal-to-noise level and
produced a new data set with a similar pixel size in the direction
of the solar X and Y coordinate.

The other two data sets were obtained on September 28, 1996 from
15:10 to 03:18 UTC on the next day in an equatorial region of the
quiet Sun, and on 16 April from 8:21 to 15:48 UTC in a south polar
region. In each of these data sets, more than 40 EUV lines with a
wide temperature coverage were included in the spectral windows to
observe almost the same region on the solar disk. However, only part
of them turned out to be strong and clean enough for our study (16
lines in the quiet Sun and 13 lines in the coronal hole). The
selected lines for the quiet Sun region are Si~{\sc{i}} (1258.77
{\AA}), O~{\sc{i}} (948.7 {\AA}), N~{\sc{ii}} (1083.98 {\AA}),
O~{\sc{ii}} (834.46 {\AA}), S~{\sc{iii}} (1077.14 {\AA}),
Si~{\sc{iii}} (1113.24 {\AA}), Si~{\sc{iv}} (1393.74 {\AA}),
N~{\sc{iii}} (991.57 {\AA}), S~{\sc{iv}} (1072.97 {\AA}),
S~{\sc{vi}} (933.41 {\AA}), O~{\sc{v}} (629.74 {\AA}), O~{\sc{vi}}
(1037.63 {\AA}), Fe~{\sc{viii}} (1062.44 {\AA}), Ne~{\sc{vii}}
(562.32 {\AA}), Ne~{\sc{viii}} (770.428 {\AA}), and Mg~{\sc{x}}
(624.97 {\AA}). The chosen lines for the coronal hole region are
Si~{\sc{i}} (1258.77 {\AA}), C~{\sc{ii}} (1037.03 {\AA}),
O~{\sc{ii}} (834.46 {\AA}), S~{\sc{iii}} (1077.14 {\AA}),
Si~{\sc{iii}} (1113.24 {\AA}), Si~{\sc{iv}} (1393.74 {\AA}),
N~{\sc{iii}} (991.57 {\AA}), S~{\sc{iv}} (1072.97 {\AA}),
S~{\sc{vi}} (933.41 {\AA}), O~{\sc{vi}} (1037.63 {\AA}),
Fe~{\sc{viii}} (1062.44 {\AA}), Ne~{\sc{vii}} (562.32 {\AA}), and
Ne~{\sc{viii}} (770.428 {\AA}). We took the values of the
wavelengths from \cite{CurdtEtal2004}. The formation temperatures
were taken from \cite{Xia2003} and the Chianti data base
\citep{DereEtal1997,LandiEtal2006} and can be found in
Fig.~\ref{fig.3} and Fig.~\ref{fig.4}. For each of these two data
sets, the observed regions are not exactly the same for different
lines, especially in the quiet Sun region for which the spatial
shift between two images can reach half of the size of the
observation region.

\begin{figure*}
\sidecaption
\includegraphics[width=13cm]{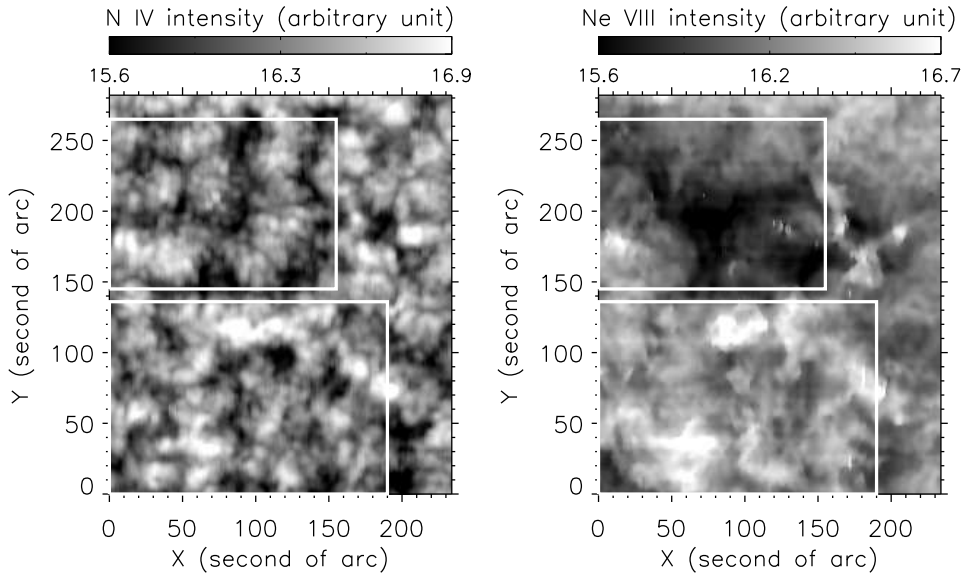}
\caption{~Intensity images of N~{\sc{iv}} (left) and Ne~{\sc{viii}}
(right) in the first data set (from 18:00 on March 7 to 17:42 on
March 8 in 1997). The intensity images are in logarithmic scale. The
two sub-regions enclosed by white lines are defined as the
coronal-hole and quiet-Sun parts and studied in detail.}
\label{fig.1}
\end{figure*}

\begin{figure*}
\sidecaption
\includegraphics[width=13cm]{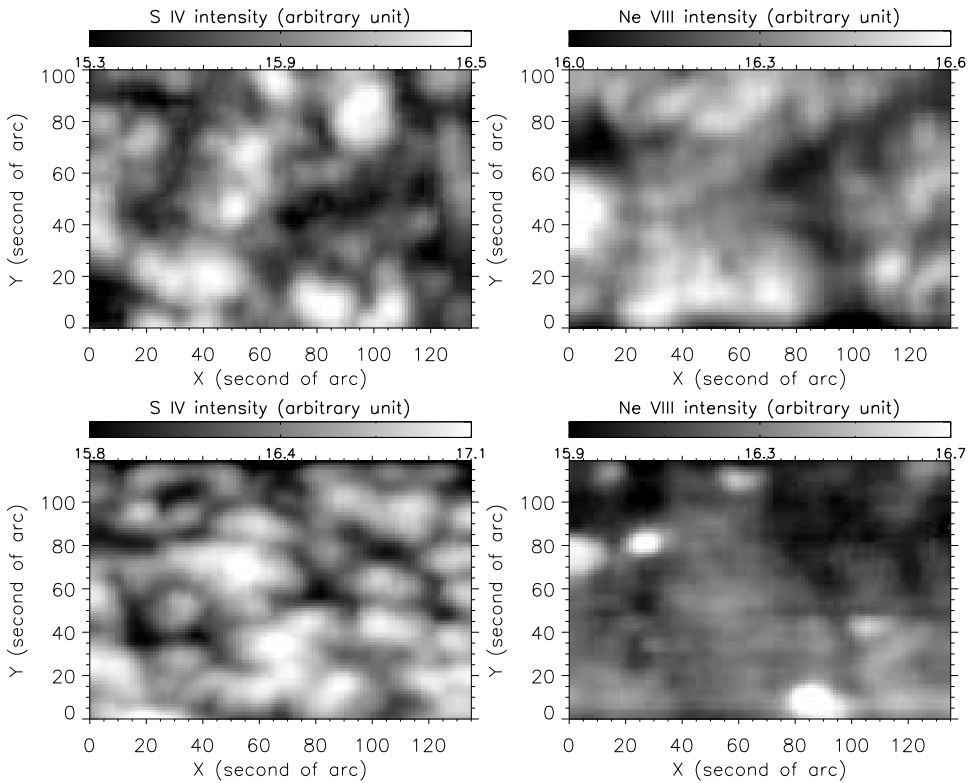}
\caption{~Intensity images of S~{\sc{iv}} (left) and Ne~{\sc{viii}}
(right) of the second data set (upper panel, observed from 15:10 on
September 28 to 03:18 on September 29) and the third data set (lower
panel, observed from 8:21 to 15:48 on April 16). The intensity
images are in logarithmic scale.} \label{fig.2}
\end{figure*}

The standard SUMER procedures for correcting and calibrating the
data were applied, including local-gain correction, flat-field
correction, geometrical-distortion and dead-time correction. A
single Gaussian fit was applied to each observed spectral line
profile. By integrating over a fitted profile, we derived the total
count rate of the line and thus obtained its intensity that was used
to build an image of each line. Fig.~\ref{fig.1} shows the intensity
images of N~{\sc{iv}} (left) and Ne~{\sc{viii}} (right) for the
first data set. The two sub-regions enclosed by the white rectangles
are defined as the coronal-hole and quiet Sun region, respectively,
which are further studied in detail. Fig.~\ref{fig.2} shows the
intensity images of S~{\sc{iv}} (left) and Ne~{\sc{viii}} (right) of
the second data set (upper frames) and of the third data set (in the
lower frames).

Since for the second and third data set the exposure times are only
30~s and 20~s, respectively, the Doppler shift and width produced by
a single Gaussian fit could not be reliably determined for many of
the spectra. So for these two data sets, we only studied the
characteristic sizes as inferred from intensity images. But for the
first data set, since the exposure time was 90~s, and as we binned
the data over four exposures, we could determine the Doppler shift
and non-thermal width of any line very well.

To determine the Doppler shift, we first estimated the instrumental
shift caused by thermal deformations of the optical system of SUMER
by using the same technique adopted in \cite{DammaschEtal1999}.
Since there is no chromospheric line in our spectral windows, we
could not carry out an absolute wavelength calibration by using a
certain cold line. Instead, we followed the method employed by
\cite{AiouazEtal2005} and obtained the dopplergram after assuming a
mean Doppler shift value for all the spectra of each line in the
first data set. We took a value of -2.5~km/s (blue shift) for
Ne~{\sc{viii}} and 7~km/s (red shift) for the other lines from
\cite{Xia2003}, in which the results were based on a statistical
study.

The measured broadening of a line profile results from a combination
of instrumental broadening, thermal and non-thermal motion
\citep{ChaeEtal1998, Xia2003}. For the line profiles recorded by
SUMER, instrumental broadening can be removed by applying the IDL
procedure $con_{-}width_{-}funct_{-}3.pro$ available in $SSW$(the
Solar Software). With the assumption of local thermal equilibrium
the corresponding thermal velocity, and thus its temperature, can be
calculated for a certain ion. After removing the instrumental and
temperature broadening, we thus obtained the image of the
non-thermal velocity for each line in the first data set.

In order to investigate the links between the 3-D magnetic
structures and the long-lasting transition-region features, we
successfully reconstructed the magnetic structure above the
photosphere by using the force-free model proposed by
\cite{Seehafer1978}, following the work of many authors
\citep{MarschEtal2004, WiegelmannEtal2005, TuEtal2005a, TuEtal2005b,
MarschEtal2006, HeEtal2007, TianEtal2007, TianEtal2008}. Here we
reconstructed a potential 3-D magnetic field, by applying the same
method and using the observed magnetograms which correspond to the
area observed by SUMER in each data set. For the first data set, a
Michelson Doppler Imager (MDI) magnetogram was chosen in the same
way as in \cite{AiouazEtal2005}. For the other two data sets, we
first obtained a full-disk Kitt-Peak magnetogram observed during the
scan period of SUMER, and then used a sub-magnetogram with a size
$40^{\prime\prime}$ larger than the SUMER observation region to
perform the field extrapolation. The line-of-sight magnetic field
components measured by the  magnetographs were then converted to the
component perpendicular to the solar surface, by simply considering
and correcting for the cosine effect. We thus calculated the
magnetic field up to 40~Mm above the photosphere.

\section{Results and discussion}

The main motivation for this work is to study the temperature or
height variation of the sizes of transition region structures in a
coronal hole and the quiet Sun. In order to estimate the
characteristic sizes of different features present in the
chromosphere, transition region and lower corona, we calculated the
ACF for each image of intensity, Doppler shift, and non-thermal
width of EUV lines obtained by SUMER, as well as the corresponding
extrapolated magnetic field at different heights. The calculation of
ACF is based on Fourier analysis and can provide a statistical
estimate of the scale length of any features in an image. We first
calculated the 2-D ACF of each image by shifting it in both
dimensions, and then took its angular average in rings of integer
radius and derived a curve of the correlation coefficient against
radial distance \citep{PatsourakosEtal1999}. The HWHM of the ACF is
considered to be the scale size of the feature, while the distance
between the primary and secondary peak corresponds to the mean
distance between two adjacent features \citep{PatsourakosEtal1999,
GontikakisEtal2003, RavindraVenk2003}. Here we are only interested
in the HWHM. The smaller the features, the sharper the central peak
of the ACF will be \citep{GontikakisEtal2003}.

\begin{figure*}
\resizebox{\hsize}{!}{\includegraphics{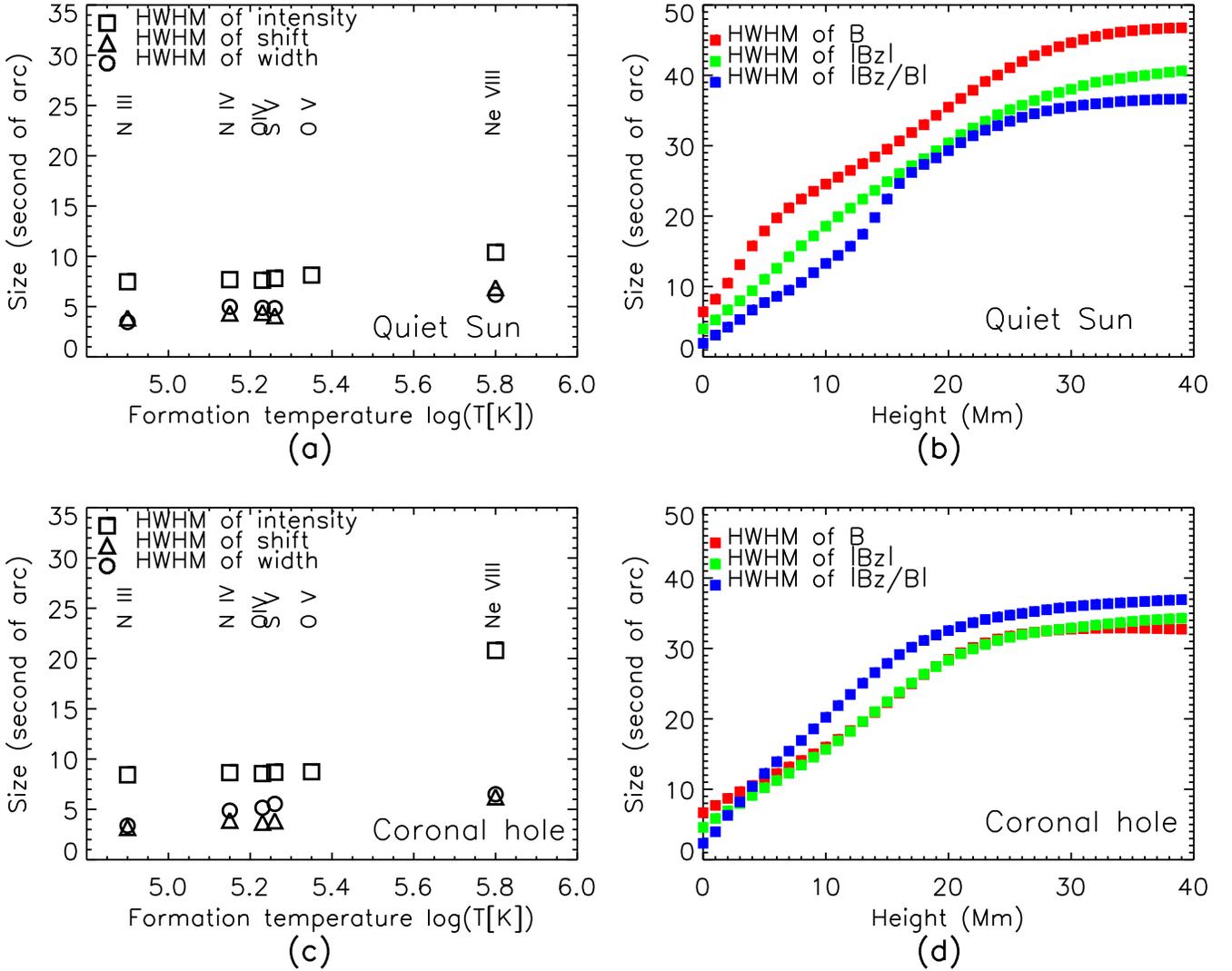}}
\caption{~Temperature/Height variations of the characteristic sizes
of intensity, Doppler shift and non-thermal width (left), as well as
of the extrapolated magnetic field (right) in the quiet Sun (upper
panel) and coronal hole (lower panel) for the first data set.}
\label{fig.3}
\end{figure*}

In Fig.~\ref{fig.3}(a) and (c), the inferred HWHM of intensity,
Doppler shift and non-thermal width are shown for all the lines
selected in the first data set for the quiet Sun and coronal hole
respectively. The O~{\sc{v}} (760.43{\AA}) line is blended with
O~{\sc{v}} (760.21{\AA}) \citep{Aiouaz2008}. The blend is weak and
the line is about 5 pixels away from O~{\sc{v}} (760.43{\AA}), so
that we can still study the intensity image. But the blend can have
a large impact on the determination of Doppler shift and non-thermal
width. Therefore, we did not study the images of the shift and width
of this line. Fig.~\ref{fig.3} demonstrates that the intensity size
is larger than that associated with the Doppler shift and
non-thermal width, a result has been found before by
\cite{GontikakisEtal2003}. It could be due to the orientation of the
magnetic field, since the observed Doppler shift and non-thermal
motion correspond to the line-of-sight components, which are
expected to be large only in the vertical legs of magnetic loops. We
found that the sizes of all three features are rather stable in the
middle transition region. The sizes of the Ne~{\sc{viii}} structures
are larger than those derived by means of middle-transition-region
lines (Log T$<$5.7), which is especially true in the coronal hole
for the size of Ne~{\sc{viii}} intensity structures. The strong
increase of the size of the Ne~{\sc{viii}} intensity features
indicates that the magnetic field expands much more strongly with
height in coronal hole than in the quiet Sun region. The field of
view selected to define the coronal hole in this data set includes a
relatively bright structure in the intensity image of
Ne~{\sc{viii}}, centered at
$(x,y)=(20^{\prime\prime},170^{\prime\prime})$ according to the axes
in Fig.~\ref{fig.1}. This bright feature was omitted from the
delimitation of the coronal hole in \cite{LemaireEtal1999} and
\cite{AiouazEtal2005}. We computed the HWHM after omitting this
structure (the field of view selected to define the coronal hole was
changed from
$(x:0^{\prime\prime}-155^{\prime\prime};y:145^{\prime\prime}-265^{\prime\prime})$
into
$(x:50^{\prime\prime}-155^{\prime\prime};y:145^{\prime\prime}-265^{\prime\prime})$).
The result reveals that except for the size of Ne~{\sc{viii}}
intensity feature, which changes from about $20^{\prime\prime}$ to
$18^{\prime\prime}$ and does not influence our result, all of the
sizes of features do not show visible changes. Thus, we believe that
the influence of the bright feature on our results can be neglected.

The two right panels of Fig.~\ref{fig.3} show the height variation
of the sizes of the corresponding (extrapolated) coronal field
parameters $B$, $\left\vert B_z \right\vert$ and $\left\vert
B_z/B\right\vert$ in the quiet Sun and coronal hole, respectively.
In the coronal hole, the HWHM of $\left\vert B_z \right\vert$ and
$B$ increase linearly with height below 10~Mm.  The sizes (in terms
of HWHM) of the two features are equal to each other at 10 Mm,
implying that small loops only reside below 10~Mm and the magnetic
field is entirely open above 10~Mm. This result is consistent with
the result of \cite{WiegelmannSolanki2004}, in which high and long
closed loops are found to be extremely rare in coronal holes. In
contrast, in the quiet Sun region the HWHM of $\left\vert B_z
\right\vert$ is always smaller than that of $B$, which means that
loops with different heights up to 40~Mm exist in the quiet Sun
region. $\left\vert B_z/B \right\vert$ is an indicator of the
inclination of a magnetic field line with respect to the horizontal
direction. So the height variation of the size of $\left\vert B_z/B
\right\vert$ can reveal the expansion of funnels. It is apparent
that below 10~Mm the size of $\left\vert B_z/B \right\vert$
increases much more strongly with height in the coronal hole than in
the quiet Sun, indicating that funnels expand more strongly with
height in the coronal hole. This is consistent with the result that
the size of the Ne~{\sc{viii}} intensity structures increases
dramatically in the coronal hole. We found that at 10~Mm, the HWHM
of $\left\vert B_z/B \right\vert$ is about $10^{\prime\prime}$ in
the quiet Sun and $20^{\prime\prime}$ in the coronal hole, which is
almost the same as the HWHM of the Ne~{\sc{viii}} intensity features
in both regions.

\begin{figure*}
\resizebox{\hsize}{!}{\includegraphics{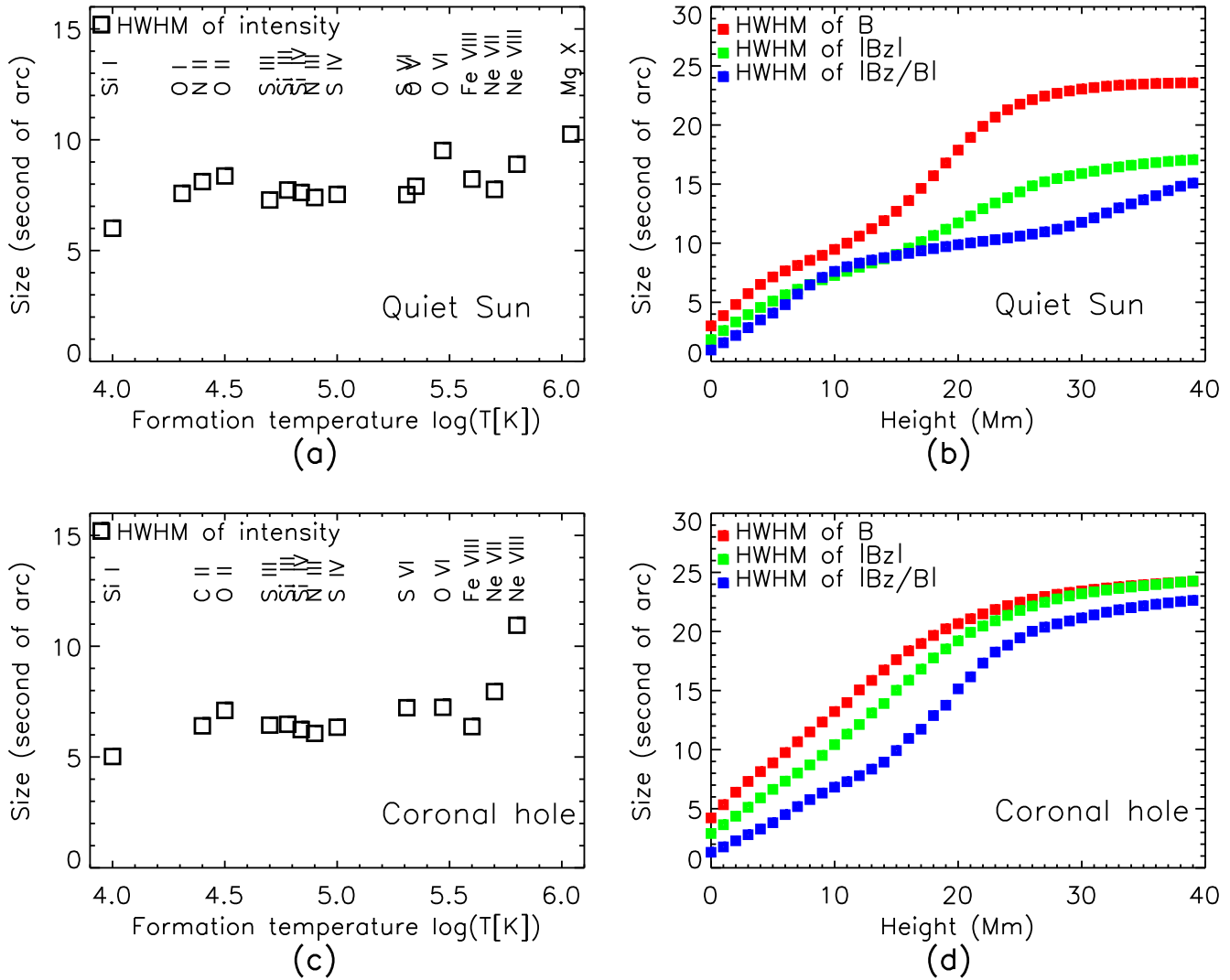}}
\caption{Temperature/height variations of the characteristic sizes
of line intensity structures (left) and of structures inferred from
the coronal magnetic field parameters (right) in the quiet Sun
(upper panel, for the second data set) and coronal hole (lower
panel, for the third data set).} \label{fig.4}
\end{figure*}

The results obtained from the second (quiet Sun) and third (coronal
hole) data set are presented in Fig.~\ref{fig.4}. The formation
temperatures of the lines selected here range from $10^4$~K to
$10^{6.02}$~K, and so we can study the variation of the size of
features in the intensity image from the chromosphere to the lower
corona, which is shown on the two left panels of the figure. If we
consider the HWHM of bright radiance features to be the size of the
network \citep{PatsourakosEtal1999}, then it is clear that the
network size increases from the chromosphere to the transition
region within the temperature range of Log $T=4.0 - 4.4$ in both the
quiet Sun and the coronal hole. In the transition region (Log $T=4.4
- 5.7$), we found that the network size almost does not change.
Beyond Log $T=5.7$, we can see an increase with temperature in
network size, more dramatic in the coronal hole than the quiet Sun
region, which is consistent with the result of the first data set.
The Mg~{\sc{x}} line in the second data set is blended with three
other lines, which contribute about 20\% to the total intensity
\citep{TeriacaEtal2002} and should not have a large influence on the
results. \cite{GontikakisEtal2003} found that the feature sizes
inferred from the middle transition region line (C~{\sc{iv}}, Log
$T=5.0$) are smaller than those from the chromospheric line
(Si~{\sc{ii}}, Log$T=4.4$). However, such a relation is not obvious
in Fig.~\ref{fig.4}.

As for the results obtained from the second and third data set,
concerning the properties of the coronal magnetic field, the height
variations of the sizes (derived from the HWHM) of features in
$\left\vert B_z \right\vert$ and $B$ generally are similar to those
found for the first data set. These results can be seen in the two
right panels of Fig.~\ref{fig.4}. One difference is the height above
which the two sizes become equal in the coronal hole, which occurs
at about 25~Mm and implies that small loops reside only below 25~Mm
in this coronal hole. The increment of the HWHM for $\left\vert
B_z/B \right\vert$ with height is almost the same below 10~Mm in
both areas, implying that the magnetic structures are of similar
size in the quiet Sun and coronal hole at low heights above the
photosphere. It is clear that from about 10~Mm to 20~Mm, the size of
$\left\vert B_z/B\right\vert$ increases much more strongly with
height in the coronal hole, a trend perhaps indicating that funnels
expand with height in the lower corona more strongly in the coronal
hole than in the quiet Sun. This conclusion is consistent with the
finding that the size of the Ne~{\sc{viii}} intensity features
dramatically increases with temperature in the coronal hole. It is
interesting to note that in both regions, the change in the HWHM of
$\left\vert B_z/B \right\vert$ from 10~Mm to 20~Mm seems to be
consistent with the change in the HWHM of the intensity with
temperature from the middle transition region to the lower corona.

Inspection of the two left panels of Fig.~\ref{fig.4} shows that the
sizes of intensity structures as seen in the different lines
originating in the middle transition region are not exactly the
same, which may reflect that the observation areas for the different
lines are not exactly the same in the second and third data set.
Another possible reason may be that the images, to which we applied
the autocorrelation technique, were not large enough in size, and
thus they perhaps do not represent an ideal sample in the
statistical sense. However, we calculated the ACF for artificial
images having different sizes but the same scale of the features,
and found that the HWHM of the ACF for an image does not depend
largely on its overall size. Thus, the overall sizes of the images
used in this paper are large enough for our study. On the other
hand, we are mainly interested in the height variation with size,
but not so much in the absolute value of the scale of a feature.
Therefore, we believe that our results are reliable for that
purpose.

One may find that the linear increase of the HWHM of $\left\vert B_z
\right\vert$ or $B$ as a function of altitude, up to 10 Mm, is not
followed by the network structures measured at temperatures lower
than the Ne~{\sc{viii}} formation temperature, as shown in
Fig.~\ref{fig.2} and Fig.~\ref{fig.4}. Especially from Log $T=4.4$
to Log $T=5.7$, the network size almost does not change. This result
seems to indicate that in the transition region, plasmas with
different temperatures can coexist at the same height
\citep{MarschEtal2006}, and the structures of their emission sources
have comparable sizes. Another possibility could be that the
emission of the solar plasma in the lower transition region is
intermittent and shows substantial temporal and spatial variability,
which could influence the computed HWHM
\citep{WilhelmEtal1998,Brkovi2003,GontikakisEtal2003}.

\section{Summary and conclusion}
For the scaling analysis, we applied the autocorrelation technique
to the images of intensity, Doppler shift, and non-thermal width of
several EUV lines measured by SUMER, as well as to the corresponding
magnetic field maps obtained by extrapolation of photospheric
magnetograms to different heights. The HWHM of the ACF is considered
to determine the characteristic size of a feature in the respective
image.

Concerning the sizes of magnetic structures in the transition
region, a different variation with height was found for the sizes
(in terms of HWHM) of features in the extrapolated field magnitude
$B$ and its component $\left\vert B_z \right\vert$. The sizes of the
two features are equal at a certain height in the coronal hole.
While in the quiet Sun region, the HWHM of $\left\vert B_z
\right\vert$ is smaller than that of $B$ at any height below 40~Mm.
This indicates that loops are much lower in the coronal hole than in
the quiet Sun region. The height variation of the HWHM of
$\left\vert B_z/B\right\vert$ suggests that an open magnetic
structure expands through the upper transition region and lower
corona more strongly in coronal holes than in the quiet Sun.
Furthermore, we found that in the upper part of the transition
region, the size of a line-intensity feature increases with
temperature more strongly in the coronal hole than in the quiet Sun.
We also corroborated a result found previously that the features in
dopplergrams and images of the non-thermal width are smaller than
those in the intensity images, a finding that may be due to the
line-of-sight effect.

\begin{acknowledgements}
We thank Dr. W. Curdt, Dr. M. S. Madjarska, Dr. L. Teriaca, and Dr.
D. Innes for their help in SUMER data analysis. We also thank the
referee Dr. Costis Gontikakis for his careful reading of the paper
and for the comments and suggestions.

H. Tian, C.-Y. Tu, and J.-S. He are supported under contracts
40574078, 40336053, 40436015, and by the Beijing Education Project
XK100010404, as well as the foundation Major Project of National
Basic Research contract 2006CB806305. H. Tian is also supported by
China Scholarship Council for his stay in the Max-Planck-Institut
f\"ur Sonnensystemforschung in Germany .

The SUMER project is financially supported by DLR, CNES, NASA, and
the ESA PRODEX programme (Swiss contribution). SUMER and MDI are
instruments on board SOHO, an ESA and NASA mission. We thank the
teams of SUMER and MDI for the spectroscopic and magnetic field
data. We also thank the NSO/Kitt Peak observatory for the use of the
magnetic field data.

\end{acknowledgements}

\end{document}